\documentclass[12pt]{article}
\usepackage{epsf}
\usepackage{color}
\usepackage{graphicx}
\usepackage{amsmath}
\usepackage{amssymb}
\usepackage{latexsym}
\usepackage{url}
\newcommand{\VTB}{V_{\mbox{{\scriptsize TB}}}}

\begin{document}
\setlength{\baselineskip}{18pt}

\begin{titlepage}
\begin{flushright}
OU-HET 715/2011 
\end{flushright}

\vspace*{1.2cm}

\begin{center}
{\Large\bf Predictions via large {\boldmath $\theta_{13}$} from cascades} 
\end{center}

\lineskip .75em
\vskip 1.5cm

\begin{center}
{\large 
Naoyuki Haba$^{a,}$\footnote[1]{E-mail:
\tt haba@phys.sci.osaka-u.ac.jp} and 
Ryo Takahashi$^{b,}$\footnote[2]{E-mail: 
\tt ryo.takahashi@mpi-hd.mpg.de}
}\\
\vspace{1cm}

$^a${\it Department of Physics, Graduate School of Science, Osaka
University, \\
Toyonaka, Osaka 560-0043, Japan}\\

$^b${\it Max-Planck-Institut f$\ddot{u}$r Kernphysik, Saupfercheckweg 1, \\ 
69117 Heidelberg, Germany}\\

\vspace*{10mm}

{\bf Abstract}\\[5mm]
{\parbox{13cm}{
We investigate a relation among neutrino observables, three mixing angles and 
two mass squared differences, from a cascade texture of neutrino mass matrix. 
We show an allowed region of the correlation by use of current data of neutrino
 oscillation experiments. The relation predicts sharp correlations among 
neutrino mixing angles as $0.315\lesssim\sin^2\theta_{12}\lesssim0.332$ and 
$0.480\lesssim\sin^2\theta_{23}\lesssim0.500$ with a large $\theta_{13}$ 
($0.03<\sin^22\theta_{13}<0.28$). These magnitudes are modified 
$0.310\lesssim\sin^2\theta_{12}\lesssim0.330$ and 
$0.540\lesssim\sin^2\theta_{23}\lesssim0.560$ when the charged lepton mass 
matrix also has the cascade form. 
}}
\end{center}

\end{titlepage}

\section{Introduction}

Current neutrino oscillation experiments suggest an existence of two large 
mixing angles among three generations in lepton sector \cite{Schwetz:2011qt}. 
It is well known that the two large mixing angles is suitably approximated by 
so-called tri-bimaximal mixing (TBM) \cite{TBM,TBM2}, 
 \begin{eqnarray}
  \VTB=
  \left(
    \begin{array}{ccc}
      2/\sqrt{6}  & 1/\sqrt{3} & 0           \\
      -1/\sqrt{6} & 1/\sqrt{3} & -1/\sqrt{2} \\
      -1/\sqrt{6} & 1/\sqrt{3} & 1/\sqrt{2}
    \end{array}
  \right), \label{VTB}
 \end{eqnarray}
which induces 
mixing angle,
 \begin{eqnarray}
  \sin^2\theta_{12}=\frac{1}{3},~~~\sin^2\theta_{23}=\frac{1}{2},~~~
  \sin\theta_{13}=0.
 \end{eqnarray} 
Such a characteristic form of mixing matrix strongly motivates a study of 
flavor structure in the lepton sector. Actually, there are a large number of 
models which try to realize the TBM based on a flavor symmetry, neutrino mass 
textures, and so on. 

In a study of suitably realization of the TBM, one peculiar relation among the 
neutrino observables, three mixing angles and two mass squared differences, has
 been proposed in \cite{Haba:2008dp}, that is 
 \begin{eqnarray}
  \frac{1}{9}\left(\sin^2\theta_{23}-\frac{1}{2}\right)
  -\frac{r}{4}\left(\sin^2\theta_{12}-\frac{1}{3}\right)
  -\frac{\sqrt{2}r}{27}\sin\theta_{13}=0, \label{rel}
 \end{eqnarray}
where $r\equiv\sqrt{\Delta m_{21}^2/\Delta m_{31}^2}$, and $\theta_{ij}$ 
$(i,j=1,2,3)$, $\Delta m_{21}^2$ and $\Delta m_{31}^2$ are the leptonic mixing 
 angles and two mass squared differences of neutrinos, respectively. Notice 
that the exact TBM satisfies this relation independently of the mass squared 
differences. Moreover, this relation also shows 
 correlations among deviations from the TBM. In fact, this attractive relation 
is derived from so-called cascade texture \cite{Haba:2008dp} with hierarchical 
neutrino masses. 

A typical cascade texture is represented by
\begin{eqnarray}
  M_{\mbox{{\scriptsize cas}}}=
   \left(
    \begin{array}{ccc}
      \delta & \delta  & \delta \\
      \delta & \lambda & \lambda \\
      \delta & \lambda & 1
    \end{array}
   \right)v~~~\mbox{ with }~~~|\delta|\ll|\lambda|\ll1,
  \label{cas}
\end{eqnarray}
where $v$ denotes an overall mass scale. In ref.~\cite{Haba:2008dp}, it has 
been pointed out that the TBM can be realized at a leading order in type-I 
seesaw mechanism \cite{seesaw}-\cite{seesaw3} if the neutrino Dirac mass matrix
 is taken as the cascade form.\footnote{There are some kinds of textures, which
 can lead to two large leptonic mixing angles with vanishing or non-vanishing 
$\theta_{13}$. For instance, the Fritzsch-type \cite{Fritzsch:1977vd} lepton 
mass matrices, which is classified to two-zero textures, can predict 
non-vanishing $\theta_{13}$ \cite{Fukugita:1992sy,Fukugita:2003tn}. One of 
interesting features of cascade texture is that it can also lead to two large 
leptonic mixing angles with either vanishing or non-vanishing $\theta_{13}$ 
even though the texture is hierarchical structure as we will show below. Such a
 hierarchical mass structure might be realized by the Froggatt-Nielsen 
mechanism \cite{Froggatt:1978nt}.} Realizations of such a cascade texture have 
also been discussed in terms of flavor symmetries and extra-dimensions 
\cite{Haba:2008dp,su5}. We call the model which induces the cascade texture 
``cascade model''. We here comment on a slightly modified cascade texture, 
called hybrid cascade texture, which is given by 
\begin{eqnarray}
  M_{\mbox{{\scriptsize hyb}}} =
  \left(
    \begin{array}{ccc}
      \epsilon & \delta'  & \delta'  \\
      \delta'  & \lambda' & \lambda' \\
      \delta'  & \lambda' & 1
    \end{array}
  \right)v',~~~\mbox{ with }~~~|\epsilon|\ll|\delta'|\ll|\lambda'|\ll1. 
  \label{hyb}
 \end{eqnarray}
This can naturally fit a quark sector, masses and mixing angles. Thus, there 
have been some researches, where the (hybrid) cascade textures can really 
reproduce the suitable masses and mixing angles of the SM fermions at a low 
energy regime in the frameworks of SUSY SU(5) \cite{su5} and SUSY SO(10) 
\cite{Adulpravitchai:2010em} GUTs. However, it should be noticed that the TBM 
in the lepton sector is hardly realized by any seesaw mechanism 
\cite{Haba:2008dp,su5,Adulpravitchai:2010em}.  

In this letter, we investigate the relation \eqref{rel} predicted from the 
original cascade model, and examine verifiability of cascade in the lepton 
sector. The latest global analyses of three-flavor neutrino parameters 
\cite{Schwetz:2011qt} give
 \begin{eqnarray}
  \sin^2\theta_{12}=0.316\pm0.016,~~~\sin^2\theta_{23}=0.51\pm0.06,~~~
  \sin\theta^2_{13}=0.017^{+0.007}_{-0.009},
 \end{eqnarray}
at $1\sigma$ level for normal neutrino mass hierarchy (NH). In addition there 
are recent observations of $\nu_\mu\rightarrow\nu_e$ oscillation by T2K 
experiment \cite{T2K}, which suggested 
 \begin{eqnarray}
  0.03<\sin^22\theta_{13}<0.28,
 \end{eqnarray}
at 90\% C.L. for NH with $\delta_{\text{CP}}=0$ \cite{Abe:2011sj}. This 
experimental result of (a non-vanishing or) large $\sin\theta_{13}$ motivates 
studies of deviation from the TBM to search a true physics which determine the 
lepton flavor structure, and screens a large number of neutrino (lepton) flavor
 models.\footnote{See e.g. \cite{Plentinger:2005kx} for early and general 
discussions of deviations from TBM (in particular, see e.g. 
\cite{Goswami:2009yy,Shimizu:2011xg} for the recent discussions of a large 
$\theta_{13}$), \cite{Shimizu:2010pg} for general discussions of deviations 
from TBM and quark-lepton complementarity \cite{Raidal:2004iw,Minakata:2004xt},
 and \cite{He:2011kn}-\cite{Araki:2011wn} for discussions of deviations from 
TBM including the latest T2K results.}

The letter is organized as follows: In section 2, we will give a brief review 
of the cascade model and investigate predictions from the model as focusing on 
the recent data of neutrino oscillation experiments. The section 3 is devoted 
to a summary.

\section{Cascade model and probing a relation among neutrino observables}

In this section, we present a brief review of the cascade texture and 
investigate predictions from it as focusing on recent data of neutrino 
oscillation experiments.

\subsection{Cascade neutrino texture}

At first, we give a brief review of the cascade neutrino texture 
\cite{Haba:2008dp}. In the cascade neutrino model, the neutrino Dirac mass 
matrix takes the following cascade form:
 \begin{eqnarray}
  M_{\nu D}=
   \left(
    \begin{array}{ccc}
      \delta & \delta   & \delta \\
      \delta & \lambda  & -\lambda \\
      \delta & -\lambda & 1
    \end{array}
   \right)v~~~\mbox{ with }~~~|\delta|\ll|\lambda|\ll1.
 \end{eqnarray}
This mass matrix can lead to experimentally favored (nearly TBM) mixing angles 
with NH in the context of type-I seesaw mechanism. Such types of mass texture 
have often been seen in the lepton mass models, e.g. with the vacuum alignments
 and non-Abelian flavor symmetry (see, for example, refs.~[7] in 
\cite{Haba:2008dp}). Mass eigenvalues of light neutrinos, $m_i$, in the model 
are given by 
 \begin{eqnarray}
  m_1 &=& \frac{v^2}{6M_3}, \label{m1} \\
  m_2 &=& \left(\frac{1}{3M_3}+\frac{3\delta^2}{M_1}\right)v^2, \\
  m_3 &=& \left(\frac{1}{2M_3}+\frac{2\lambda^2}{M_2}\right)v^2, \label{m3}
 \end{eqnarray}
in the diagonal basis of right-handed neutrino mass matrix, 
$M_R=$Diag$[M_1,M_2,M_3]$. The cascade neutrino model leads to the NH in order 
to realize the tri-bimaximal mixing at the leading order \cite{Haba:2008dp}. 
Thus, we perturbatively computed to give eqs.~\eqref{m1}-\eqref{m3} around 
$m_1/m_{2,3}$ and $\delta/\lambda$, which are small quantities to be 
consistent with experimental values. In the same perturbation, the mixing angle
 are evaluated as
 \begin{eqnarray}
  \sin^2\theta_{12} &=& \left|\frac{1}{\sqrt{3}}
                        -\frac{2}{\sqrt{3}}\frac{m_1}{m_2}\right|^2, 
                        \label{12}\\
  \sin^2\theta_{23} &=& \left|-\frac{1}{\sqrt{2}}
                        +\frac{1}{\sqrt{2}}\frac{m_1(m_3-m_2)}{m_3(m_3-m_2)}
                        +\frac{\delta}{3\sqrt{2}\lambda}\frac{m_2}{m_3-m_2}
                        \right|^2, \\
  \sin^2\theta_{13} &=& \left|\frac{\delta}{\sqrt{2}\lambda}
                              \frac{m_3-2m_2/3}{m_3-m_2}
                        +\frac{\sqrt{2}m_1m_2}{m_3(m_3-m_2)}\right|^2, 
                        \label{13}
 \end{eqnarray}
in the diagonal basis of the charged lepton mass matrix. Notice again that this
 cascade neutrino model leads to the TBM at leading order. In other words, the 
corrections of next-leading order shift the mixing angles form the exact TBM. 
We will focus on this point in the following subsections.

It can be seen that there are four combinations of independent model 
parameters, $m_i$ and $\delta/\lambda$, while the five observables exist 
(three mixing angles and two mass squared differences can be expressed by $m_i$
 and $\delta/\lambda$). Therefore, one {\it parameter independent relation} 
among neutrino observables must exist, that is just \eqref{rel}, with real 
parameters in the model. 

\subsection{Probing a relation among neutrino observables}

Now let us investigate a predicted relation \eqref{rel} from the above cascade 
model to examine the verifiability of the model through the data of neutrino 
oscillation experiments.

\begin{figure}
\hspace{4cm}(a)\hspace{7.8cm}(b)

\includegraphics[scale=0.95]{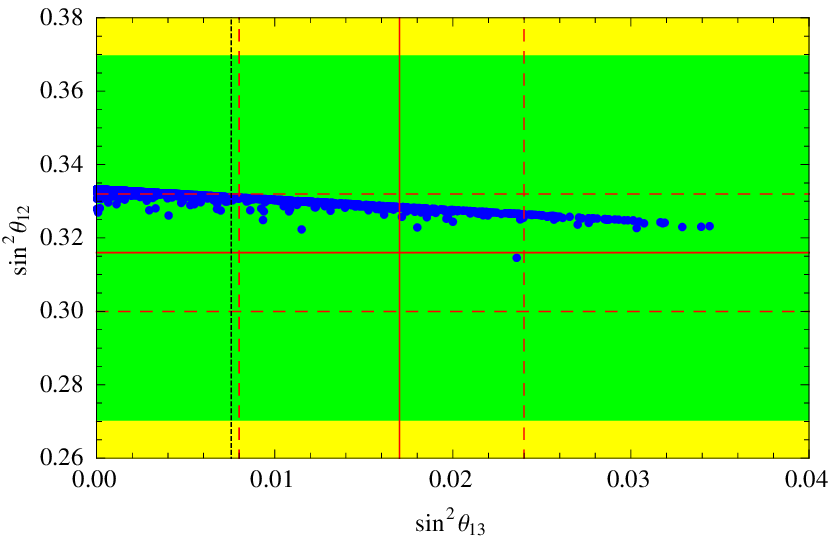}
\includegraphics[scale=0.95]{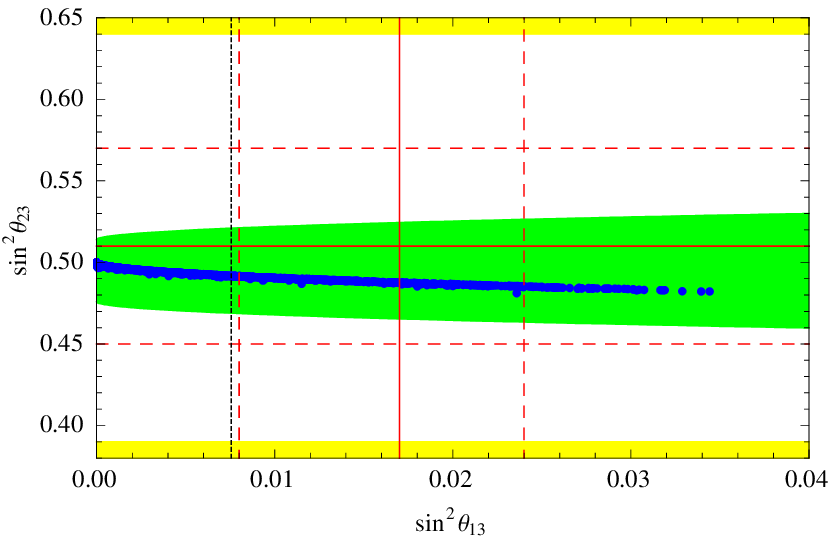}
\begin{center}
\hspace{0.5cm}(c)

\includegraphics[scale=0.95]{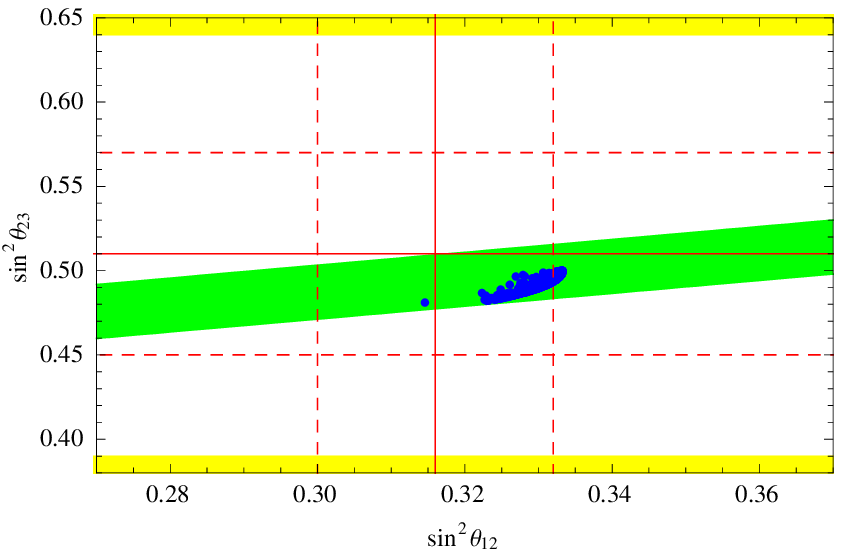}
\end{center}
\caption{Correlations among mixing angles in the cascade model of neutrino mass
 matrix with $\delta_{\text{CP}}=0$.}
\label{fig1}
\end{figure}

We give numerical plots in Fig.~\ref{fig1}.
These figures show predicted regions from a relation among neutrino observables
 in the cascade model. This numerical simulation is based on random plots 
including a mild hierarchical cascade, $0<\delta<\lambda<0.1$. Therefore, this 
simulation could scan all classes of cascade model, that is, from a mild 
hierarchy up to a rapid one. From this complete scan, about 1600 viable models 
have been chosen among 10000 models. Too mild hierarchical models are 
automatically screened by experimental data. The Fig.~\ref{fig1} (a), (b) and 
(c) are drown in $(\sin^2\theta_{13},\sin^2\theta_{12})$, 
$(\sin^2\theta_{13},\sin^2\theta_{23})$ and 
$(\sin^2\theta_{12},\sin^2\theta_{23})$ planes, respectively. The upper and 
lower flat (yellow) shaded regions in all figures indicate regions out of 
$3\sigma$ level for the vertical axes. The horizontal regions in all figures 
are shown within the $3\sigma$ levels. The (red) lighter solid and dashed lines
 correspond to the best fit and $1\sigma$ lines for all mixing angles. The 
(black) darker solid lines in the Fig.~\ref{fig1} (a) and (b) are the lower 
bound ($\sin^22\theta_{13}=0.03$) at 90\% C.L. for NH with 
$\delta_{\text{CP}}=0$ as reported by the latest T2K experiment (the upper 
bound, $\sin^22\theta_{13}=0.28$ is out of the figure). The (green) darker 
regions show where the \eqref{rel} is satisfied with each value of three mixing
 angles in $3\sigma$ ranges and the best fit values of two mass squared 
differences. The (blue) plots are the predicted points from the cascade model. 
Note that since predicted mixing angles from the cascade model are strongly 
correlated each other as shown in \eqref{12}-\eqref{13}, the (blue) plots are 
on the partial regions of the (green) darker ones, which are covered by all 
$3\sigma$ data without correlations of the mixing angles. 

\begin{figure}[ht]
\hspace{4cm}(a)\hspace{7.8cm}(b)

\includegraphics[scale=0.95]{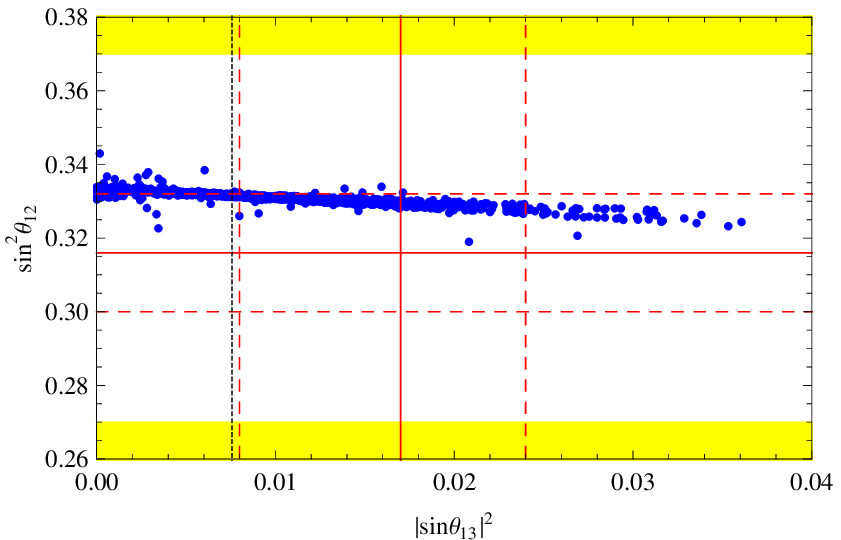}
\includegraphics[scale=0.95]{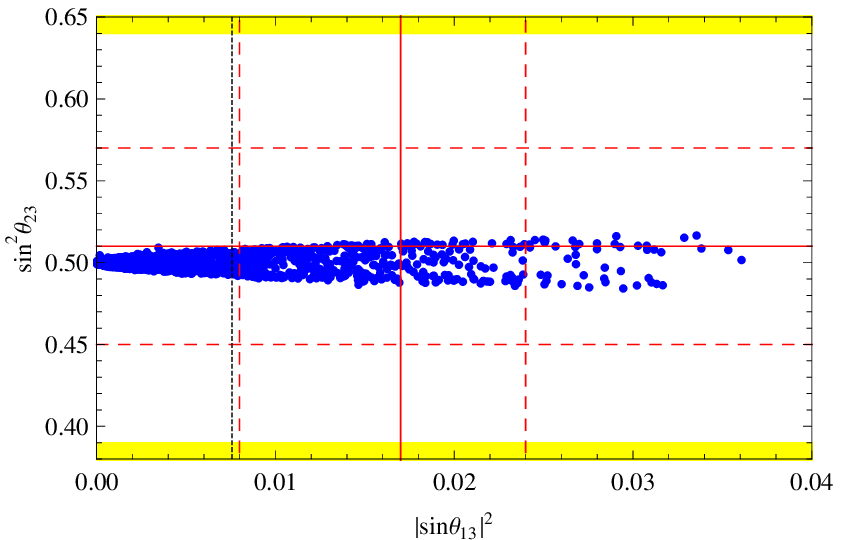}
\begin{center}
\hspace{0.5cm}(c)

\includegraphics[scale=0.95]{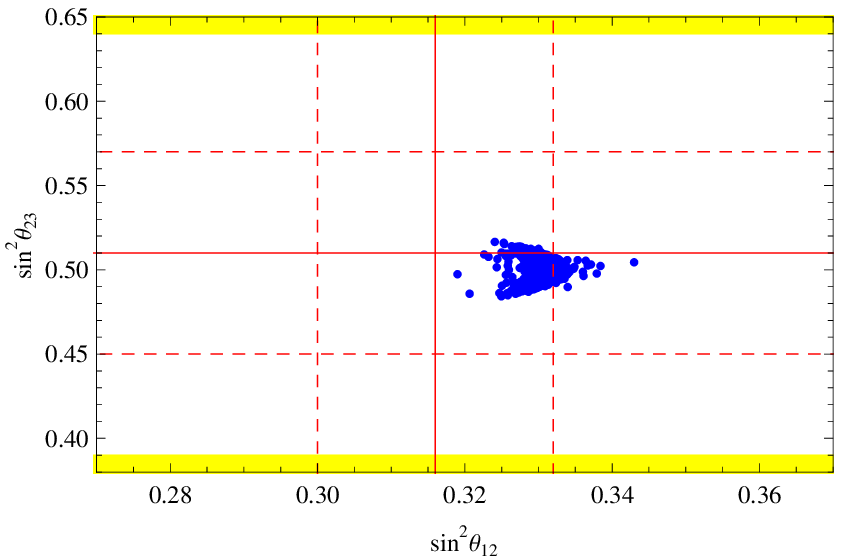}
\end{center}
\caption{Correlations among mixing angles in the cascade model of neutrino mass
 matrix with $\delta_{\text{CP}}\neq0$.}
\label{fig1-1}
\end{figure}

We can see that there are relatively strong correlations among each mixing 
angle compared with other neutrino flavor models. This is one of advantages of 
the cascade neutrino model to check the model. In particular, we can predict 
$0.315\lesssim\sin^2\theta_{12}\lesssim0.332$ and 
$0.480\lesssim\sin^2\theta_{23}\lesssim0.500$ with a relatively large 
$\theta_{13}$, e.g. $\sin^2\theta_{13}\sim0.008$, which corresponds to the 
lower bound from T2K. The above computation has been done with the vanishing 
Dirac CP phase, $\delta_{\text{CP}}$. When $\delta_{\text{CP}}\neq0$, the above
 correlations slightly weaken but the predictions of mixing angles do not 
change drastically as $0.320\lesssim\sin^2\theta_{12}\lesssim0.333$ and 
$0.480\lesssim\sin^2\theta_{23}\lesssim0.510$ for 
$0.008\lesssim|\sin\theta_{13}|^2$ which are shown in Fig.~\ref{fig1-1}. 
Note that only scatter plots are shown in Fig.~\ref{fig1-1} with 
$\delta_{\text{CP}}\neq0$ case, since the relation (\ref{rel}) is established 
only for real parameters. Anyhow, we emphasize that the cascade model is surely
 predictive, and thus it would be checked by the T2K and other future neutrino 
experiments such as the Double Chooz \cite{Ardellier:2006mn} (whose future 
sensitivity is $\sin^2\theta_{13}=0.07$ at 2011$\sim$ year), RENO 
\cite{:2010vy} ($\sin^2\theta_{13}=0.03$ at 2011$\sim$ year) and Daya-Bay 
\cite{Guo:2007ug} ($\sin^2\theta_{13}=0.01$ at 2012$\sim$ year) collaborations 
\cite{Mezzetto:2010zi} thanks to the above strong correlations among mixing 
angles.

\subsection{Cascade charged lepton mass texture}

It might be more natural that the charged lepton mass matrix also takes the 
cascade form in the sense that the neutrino Dirac mass matrix of the cascade 
form is obtained from an $U(1)$ flavor symmetry and/or other 
dynamics~\cite{Haba:2008dp}. Thus, here we research the case when the charged 
lepton mass matrix also has the cascade form,  
 \begin{eqnarray}
  M_e=
   \left(
    \begin{array}{ccc}
      \delta_e & \delta_e  & \delta_e  \\
      \delta_e & \lambda_e & \lambda_e \\
      \delta_e & \lambda_e & 1
    \end{array}
   \right)v~~~\mbox{ with }~~~|\delta_e|\ll|\lambda_e|\ll1. \label{charged}
 \end{eqnarray}
Note that the magnitudes of cascade parameters, $\delta_e$ and $\lambda_e$, 
should be evaluated from the experimentally observed values of charged lepton 
masses, $m_e$, $m_\mu$ and $m_\tau$, which are given by
 \begin{align}
  |\lambda_e|&\simeq\frac{m_\nu}{m_\tau}\simeq6\times10^{-2}, \label{lam}\\
  |\delta_e|&\simeq\frac{m_e}{m_\tau}\simeq3\times10^{-4}.    \label{del}
 \end{align}
It can be found from \eqref{charged}-\eqref{del} that the contributions to the 
mixing angles from the charged lepton sector are small. Therefore, the total 
lepton mixing angles can be estimated at the first order perturbation as
 \begin{eqnarray}
  \sin^2\theta_{12} &=& \left|\frac{1}{\sqrt{3}}
                        -\frac{2}{\sqrt{3}}\frac{m_1}{m_2}
                        -\frac{1}{\sqrt{3}}\frac{m_e}{m_\mu}\right|^2,  \\
  \sin^2\theta_{23} &=& \left|-\frac{1}{\sqrt{2}}
                        +\frac{1}{\sqrt{2}}\frac{m_1(m_3-m_2)}{m_3(m_3-m_2)}
                        +\frac{\delta}{3\sqrt{2}\lambda}\frac{m_2}{m_3-m_2}
                        -\frac{1}{\sqrt{2}}\frac{m_\mu}{m_\tau}
                        \right|^2, \\
  \sin^2\theta_{13} &=& \left|\frac{\delta}{\sqrt{2}\lambda}
                              \frac{m_3-2m_2/3}{m_3-m_2}
                        +\frac{\sqrt{2}m_1m_2}{m_3(m_3-m_2)}
                        +\frac{1}{\sqrt{2}}\frac{m_e}{m_\mu}\right|^2. 
 \end{eqnarray}
One can see that the solar neutrino mixing is little affected, on the other 
hand, as for the atmospheric neutrino mixing, the charged lepton effect often 
dominates. And the magnitude of the contribution to the reactor neutrino mixing
 is of negligible order. Since the hierarchy in the charged lepton mass matrix 
can be expressed by the observables as \eqref{lam} and \eqref{del}, the strong 
correlation among neutrino observables still holds but (\ref{rel}) is slightly 
modified as
 \begin{eqnarray}
  \frac{1}{9}\left(\sin^2\theta_{23}-\frac{1}{2}-\frac{m_\mu}{m_\tau}\right)
  -\frac{r}{4}\left(\sin^2\theta_{12}-\frac{1}{3}\right)
  -\frac{\sqrt{2}r}{27}\sin\theta_{13}=0, \label{relc}
 \end{eqnarray}
by including the charged lepton contributions in the first order approximation.
  
\begin{figure}
\hspace{4cm}(a)\hspace{7.8cm}(b)

\includegraphics[scale=0.95]{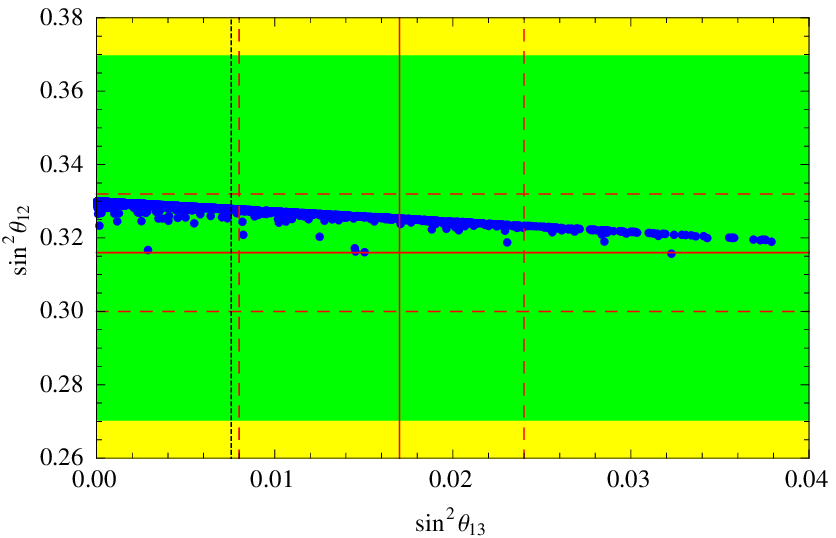}
\includegraphics[scale=0.95]{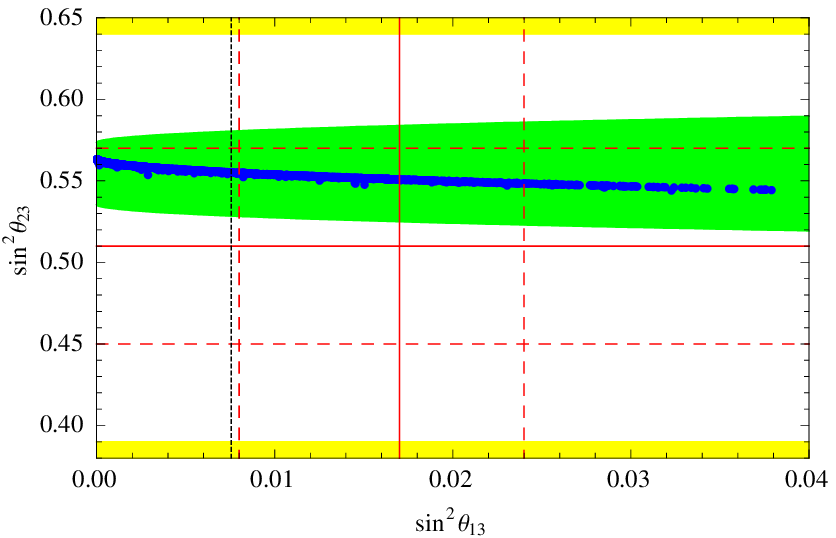}
\begin{center}
\hspace{0.5cm}(c)

\includegraphics[scale=0.95]{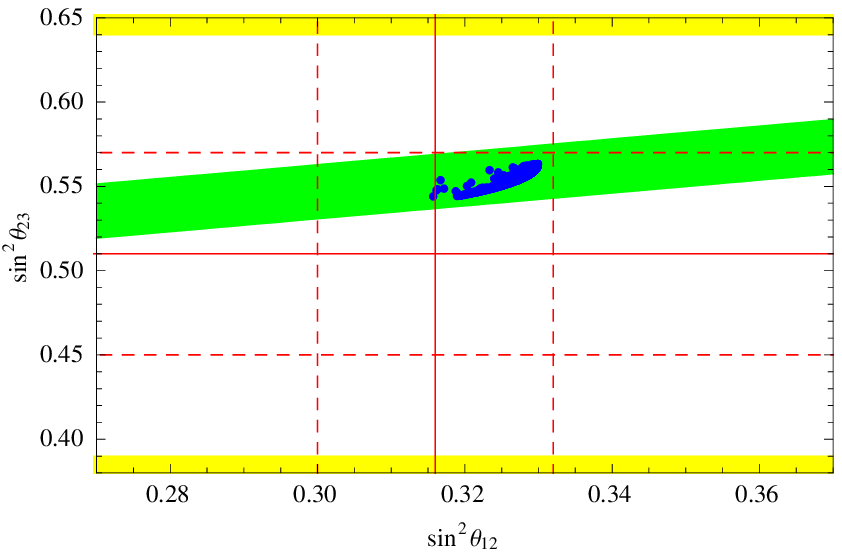}
\end{center}
\caption{Correlations among mixing angles in the cascade model of the lepton 
sector with $\delta_{\text{CP}}=0$.}
\label{fig2}
\end{figure}

\begin{figure}
\hspace{4cm}(a)\hspace{7.8cm}(b)

\includegraphics[scale=0.95]{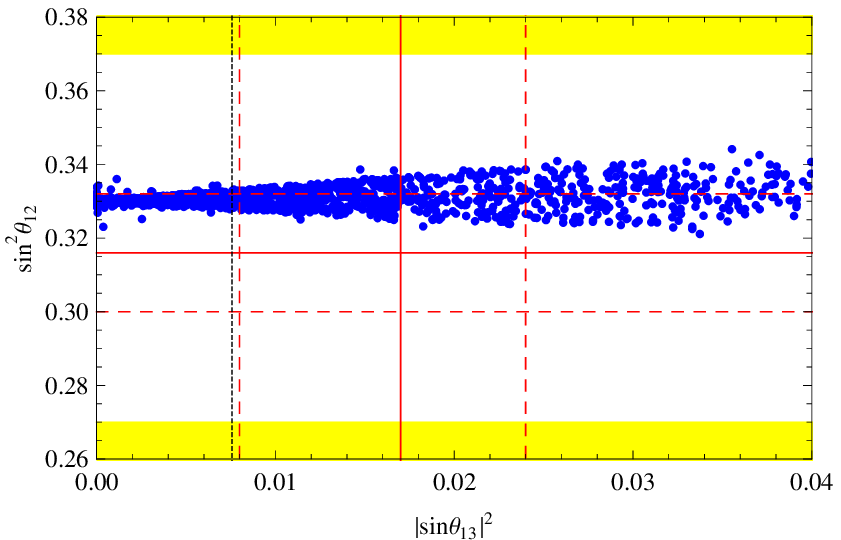}
\includegraphics[scale=0.95]{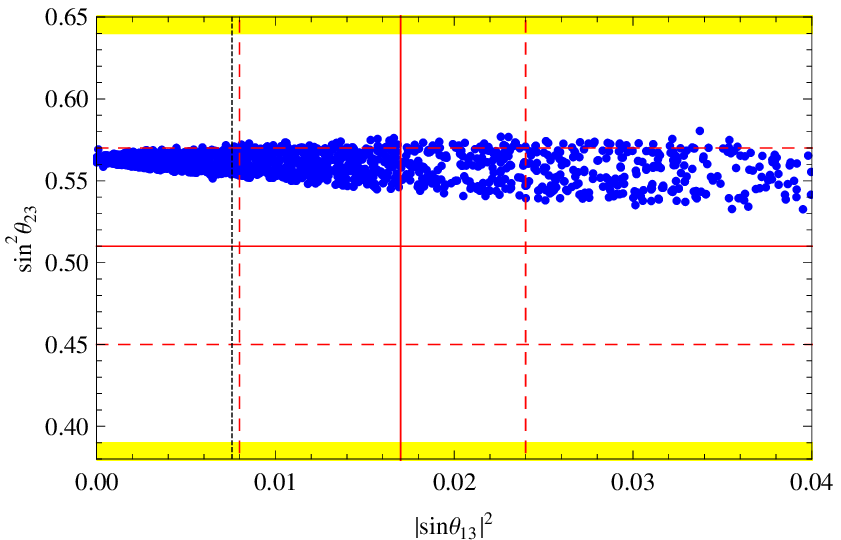}
\begin{center}
\hspace{0.5cm}(c)

\includegraphics[scale=0.95]{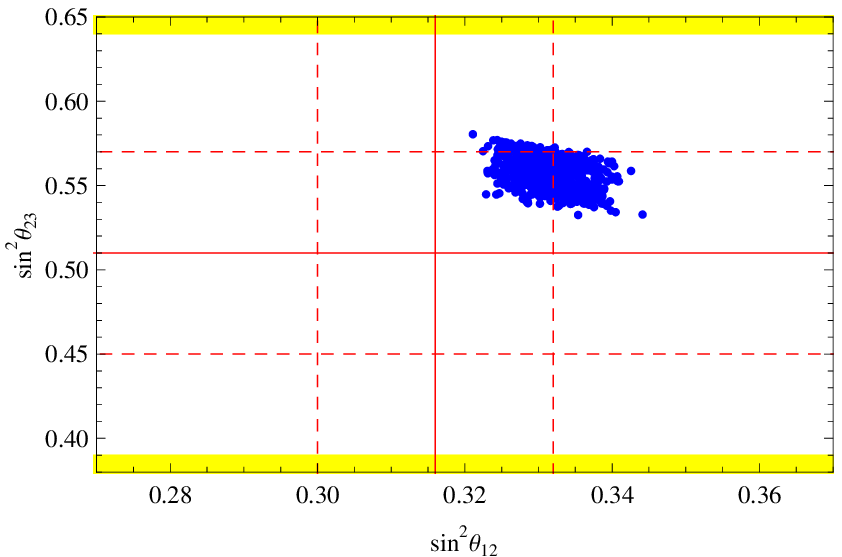}
\end{center}
\caption{Correlations among mixing angles in the cascade model of the lepton 
sector with $\delta_{\text{CP}}\neq0$.}
\label{fig2-1}
\end{figure}

We show numerical plots with the relation \eqref{relc} in Fig.~\ref{fig2}.
A fundamental setup of the numerical simulation is the same as one in the 
previous subsection. The cascade parameters in the charged lepton mass matrix 
are determined by the experimentally observed charged lepton masses. About 1600
 possible models have been also chosen among 10000 complete sets of model. This
 means that the contributions from the charged lepton mass matrix of the 
cascade form do not drastically change the results from models with the 
diagonal charged lepton mass matrix in the previous subsection. It can be seen 
that the prediction of the value of $\sin^2\theta_{12}$ with a large 
$\theta_{13}$ becomes slightly smaller compared to the case of diagonal charged
 lepton mass matrix, as $0.310\lesssim\sin^2\theta_{12}\lesssim0.330$. On the 
other hand, $\sin^2\theta_{23}$ becomes larger as 
$0.540\lesssim\sin^2\theta_{23}\lesssim0.560$. In the case of 
$\delta_{\text{CP}}\neq0$, the correlations slightly weaken and mixing angles 
are predicted as $0.320\lesssim\sin^2\theta_{12}\lesssim0.345$ and 
$0.530\lesssim\sin^2\theta_{23}\lesssim0.580$ for 
$0.008\lesssim|\sin\theta_{13}|^2$, which are shown in Fig~\ref{fig2-1}. 
Therefore, we conclude that all leptonic Dirac mass textures of the cascade 
model, also predict explicit deviations from the exact TBM, and the deviations 
are strongly correlated with each other. These would be also checked in the 
future neutrino oscillation experiments with higher sensitivities.

\section{Summary}

We have studied a relation among neutrino observables, which are three mixing 
angles and two mass squared differences. This relation is predicted from a 
cascade texture with hierarchical neutrino masses. The neutrino cascade model 
is favored by the current neutrino oscillation experiments and is supported by 
theoretical studies of new physics such as the realizations from flavor 
symmetry, extra-dimensional theory, and embedding the model into GUTs. Since 
recent data of the neutrino oscillation experiments including the latest T2K 
result might suggest the deviations from the exact TBM, we have motivated for 
the precise investigations of the relation. The relation gives strong 
correlations among each deviation of leptonic mixing angle from the TBM.

We have numerically shown predicted regions of the relation and scatter plots 
from a cascade model by use of recent data of neutrino oscillation experiments 
in two cases. One is the model with the diagonal charged lepton mass matrix, 
and the other is the case of cascade form of charged lepton mass matrix also. 
In both cases, we can see that predictions of the cascade model and deviations 
from the TBM are strongly correlated among three lepton mixing angles. This is 
a strong advantage of the cascade model for the verifiability of the model 
compared with other neutrino flavor models. We have predicted 
$0.315\lesssim\sin^2\theta_{12}\lesssim0.332$, 
$0.480\lesssim\sin^2\theta_{23}\lesssim0.500$ in the case of the diagonal 
charged lepton mass matrix, and $0.310\lesssim\sin^2\theta_{12}\lesssim0.330$, 
$0.540\lesssim\sin^2\theta_{23}\lesssim0.560$ in the case of the cascade 
charged lepton mass matrix. Hence, we conclude that the cascade model has 
predicted deviations of all mixing angles from the exact TBM with a relatively 
large $\theta_{13}$. These would be also checked in the future neutrino 
oscillation experiments with higher sensitivities.

At the end of this letter, we comment on  recent MINOS result, where 
$\theta_{13}$ can be still consistent with zero \cite{minos}. 
Figures~\ref{fig1}-\ref{fig2-1} suggest the correlations among $\theta_{ij}$ 
even if $\theta_{13}=0$, where the cascade predictions are slightly modified as
 $0.315\lesssim\sin^2\theta_{12}\lesssim0.335$. Anyhow, the cascade model 
predicts the strong correlation in wide range of $\theta_{13}$ (as 
$\sin^22\theta_{13}<0.28$).

\subsection*{Acknowledgments}

This work is partially supported by Scientific Grant by Ministry of Education 
and Science, Nos.\ 20540272, 20039006, 20025004 (NH). The work of RT is 
supported by the DFG-SFB TR 27. 


\end{document}